\documentclass[sigconf]{acmart}

\AtBeginDocument{%
  }

\usepackage{graphicx}
\usepackage{tabularx}
\usepackage{enumitem}
\usepackage{siunitx} 
\usepackage{pifont}
\usepackage[font=small,skip=0pt]{caption}
\usepackage{ifthen}

\usepackage{array} %
\newcolumntype{P}[1]{>{\centering\arraybackslash}p{#1}} %
\usepackage{subcaption} 
\usepackage{comment}
\usepackage{xspace}

\usepackage{amsmath}

\usepackage{multirow}
\usepackage{makecell}

\usepackage{booktabs}
\usepackage{threeparttable} 
\usepackage[most]{tcolorbox}
\usepackage{listings}

\lstdefinestyle{mystyle}{
    basicstyle=\ttfamily\small, 
    breakatwhitespace=false,
    breaklines=true,
    captionpos=b, 
    frame=tb, 
    framesep=2pt,
    framerule=0pt,
    aboveskip=0pt,
    belowskip=0pt,
    xleftmargin=0pt, 
    xrightmargin=0pt, 
    numbers=none, 
}

\lstdefinestyle{javaStyle}{
  language=Java,
  basicstyle=\ttfamily\scriptsize,
  keywordstyle=\color{blue!70!black}\bfseries,
  stringstyle=\color{green!50!black},
  commentstyle=\color{gray}\itshape,
  numbers=none,
  frame=single,
  framerule=0.5pt,
  rulecolor=\color{black},
  breaklines=true,
  tabsize=2,
  showstringspaces=false
}

\usepackage{xcolor}
\usepackage{etoolbox}
\usepackage{balance}

\usepackage{hyperref}
\hypersetup{
  colorlinks=true,
  linkcolor=transparent,
  anchorcolor=transparent,
  citecolor=black,
  urlcolor=black
}

\usepackage{cleveref}

\definecolor{mylightgray}{RGB}{224,224,224}

\newboolean{showcomments}
\setboolean{showcomments}{true}
\ifthenelse{\boolean{showcomments}}
 { \newcommand{\mynote}[2]{
      \fbox{\bfseries\sffamily\scriptsize#1}
        {\small$\blacktriangleright$\textsf{\emph{#2}}$\blacktriangleleft$}}}
        { \newcommand{\mynote}[2]{}}

\begin{document}


\title{Beyond Surface Similarity: Evaluating LLM-Based Test Refactorings with Structural and Semantic Awareness }

\author{Wendkûuni C. Ouédraogo}
\email{wendkuuni.ouedraogo@uni.lu}
\affiliation{
  \institution{University of Luxembourg}
 	\country{Luxembourg}
}

\author{Yinghua Li}\authornote{Corresponding author.}
\email{yinghua.li@uni.lu}
\affiliation{
  \institution{University of Luxembourg}
 	\country{Luxembourg}
}

\author{Xueqi Dang}
\email{xueqi.dang@uni.lu}
\affiliation{
  \institution{University of Luxembourg}
 	\country{Luxembourg}
}

\author{Xin Zhou}
\email{xinzhou.2020@phdcs.smu.edu.sg}
\affiliation{
  \institution{Singapore Management University}
  \country{Singapore}
}

\author{Anil Koyuncu}
\email{anil.koyuncu@cs.bilkent.edu.tr}
\affiliation{
  \institution{Bilkent University}
 	\country{Turkey}
}

\author{Jacques Klein}
\email{jacques.klein@uni.lu}
\affiliation{
  \institution{University of Luxembourg}
  \country{Luxembourg}
}

\author{David Lo}
\email{davidlo@smu.edu.sg}
\affiliation{
  \institution{Singapore Management University}
  \country{Singapore}
}

\author{Tegawend\'e F. Bissyand\'e}
\email{tegawende.bissyande@uni.lu}
\affiliation{
  \institution{University of Luxembourg}
 	\country{Luxembourg}
}

\renewcommand{\shortauthors}{Ouédraogo and Li et al.}
\begin{abstract}
\noindent
Large Language Models (LLMs) are increasingly used to refactor unit tests, improving readability and structure while preserving behavior. Evaluating such refactorings, however, remains difficult: metrics like CodeBLEU penalize beneficial renamings and edits, while semantic similarities overlook readability and modularity. We propose \textbf{CTSES}, a first step toward human-aligned evaluation of refactored tests. \texttt{CTSES} combines CodeBLEU, METEOR, and ROUGE-L into a composite score that balances semantics, lexical clarity, and structural alignment. Evaluated on 5{,}000+ refactorings from Defects4J and SF110 (GPT-4o and Mistral-Large), \texttt{CTSES} reduces false negatives and provides more interpretable signals than individual metrics. 
Our emerging results illustrate that \texttt{CTSES} offers a proof-of-concept for composite approaches, showing their promise in bridging automated metrics and developer judgments.
\end{abstract}

\ccsdesc[500]{Software and its engineering~Software testing and debugging}
\ccsdesc[300]{Computer systems organization~Neural networks}

\keywords{Test refactoring, Large Language Models, CodeBLEU, ROUGE-L, METEOR, Test evaluation, Similarity metrics, Software testing, Empirical software engineering}

\setcopyright{none}

\maketitle

\section{Introduction}
\label{intro}

LLMs are increasingly used to generate and refactor unit tests. While generation focuses on producing executable tests, refactoring aims to enhance readability and structure while preserving behavior. Yet, evaluating refactored tests remains challenging. CodeBLEU~\cite{ren2020codebleu}, widely adopted in test generation~\cite{shin2024domain} and even refactoring~\cite{deljouyi2024leveraging}, penalizes beneficial edits such as renaming, modularization, or added comments, as it emphasizes surface-level similarity rather than clarity.

We introduce \textbf{CTSES}, a first step toward human-aligned evaluation of refactored tests. \texttt{CTSES} combines CodeBLEU, METEOR~\cite{banerjee2005meteor}, and ROUGE-L~\cite{lin2004rouge,lin2004automatic} into a composite score balancing semantics, lexical clarity, and structural alignment. Unlike single metrics, \texttt{CTSES} rewards improvements in naming, modularity, and readability while safeguarding functional intent.

We evaluate \texttt{CTSES} on over \textbf{5,000 refactorings} produced by \textbf{GPT-4o}~\cite{hurst2024gpt} and \textbf{Mistral-Large-2407}\footnote{\url{https://mistral.ai/news/mistral-large-2407}} across two established benchmarks: \textbf{Defects4J}~\cite{just2014defects4j} and \textbf{SF110}~\cite{panichella2017automated}. Results show that \texttt{CTSES}: (i) aligns more closely with human judgment, (ii) reduces false negatives where CodeBLEU underrates valid refactorings, and (iii) offers tunable weighting profiles for different evaluation goals.

All code, data, and artifacts are released in our replication package.\footnote{\url{https://anonymous.4open.science/r/CTSES-75D4}}
\section{Motivating Example}
\label{motivating_example}
To illustrate CodeBLEU’s misalignment, we contrast an EvoSuite test from SF110~\cite{panichella2017automated} with a GPT-4o refactoring using a Chain-of-Thought prompt~\cite{wei2022chain}. The original test for \texttt{MacawWorkBench} (project \texttt{69\_lhamacaw}\footnote{Project identifier in SF110}) shows typical EvoSuite issues: synthetic method names (\texttt{test0}), generic variables (\texttt{stringArray0}), and no semantic cues or documentation.


\vspace{2mm}
\begin{lstlisting}[style=javaStyle, caption={Excerpt of original EvoSuite-generated test}, label=lst:evosuite_test]
@Test(timeout = 4000)
public void test0() throws Throwable {
  String[] stringArray0 = new String[2];
  stringArray0[0] = ""; stringArray0[1] = "";
  try {
    MacawWorkBench.main(stringArray0);
    fail("Expecting exception: NoClassDefFoundError");
  } catch (NoClassDefFoundError e) {
    verifyException("macaw.businessLayer.User", e);
  }
}
\end{lstlisting}

In contrast, Listing~\ref{lst:llm_test} presents the refactored version by GPT-4o. The new test improves method and variable naming, adds structured comments, and preserves all functional behavior. This reflects typical refactoring goals: improving clarity and intent without altering execution.

\vspace{2mm}
\begin{lstlisting}[style=javaStyle, caption={LLM-refactored test (GPT-4o)}, label=lst:llm_test]
@Test(timeout = 4000)
public void testMainMethodThrowsNoClassDefFoundError() throws Throwable {
    // Given: A specific array of strings to simulate command line arguments
    String[] stringArray = TEST_STRING_ARRAY.clone();
    // When: The main method of MacawWorkBench is called
    // Then: Expect a NoClassDefFoundError due to missing class initialization
    try {
        MacawWorkBench.main(stringArray);
        fail("Expecting exception: NoClassDefFoundError");
    } catch (NoClassDefFoundError e) {
        // Verify the exception is related to class initialization
        // verifyException("macaw.businessLayer.User", e);
    }
}
\end{lstlisting}

We evaluated both versions using several commonly used similarity metrics and pre-trained embedding models. Table~\ref{tab:refactor_metrics} presents the scores.

\begin{table}[ht]
\centering
\caption{Similarity scores between EvoSuite and LLM-refactored test}
\label{tab:refactor_metrics}
\scalebox{0.8}{
\begin{tabular}{lr}
\textbf{Metric} & \textbf{Score} \\
\hline
CodeBLEU & 0.433 \\
METEOR & 0.654 \\
ROUGE-L & 0.568 \\
Cosine Similarity (CodeBERT) & 0.9988 \\
Cosine Similarity (GraphCodeBERT) & 0.9963 \\
Cosine Similarity (OpenAI \texttt{text-embedding-3-small}) & 0.9669 \\
\end{tabular}
}
\end{table}

Despite equivalent behavior, the two tests differ in naming, structure, and clarity. METEOR~\cite{banerjee2005meteor} and ROUGE-L~\cite{lin2004rouge, lin2004automatic} suggest moderate similarity, while CodeBLEU remains low due to renaming. In contrast, embeddings from CodeBERT~\cite{feng2020codebert}, GraphCodeBERTGraphCodeBERT~\cite{guo2020graphcodebert}, and OpenAI’s \texttt{text-embedding-3-small}\footnote{\url{https://openai.com/index/new-embedding-models-and-api-updates/}} yield cosine scores above 0.96, confirming semantic equivalence. Among all metrics, CodeBLEU shows the greatest misalignment—despite its common use in test evaluation.

\subsection{Why CodeBLEU Falls Short}


CodeBLEU, designed for functional code, mixes lexical, syntactic, and data-flow features but is ill-suited for test refactoring. It penalizes harmless renamings (\emph{lexical bias}), ignores preserved behavior such as assertions (\emph{semantic blindness}), and treats readability gains like clearer naming or modularization as noise (\emph{structural penalization}). Prior studies confirm these flaws: Shin et al.~\cite{shin2024domain} highlight sensitivity to identifiers, and Nie et al.~\cite{nie2023learning} show valid compilable tests being penalized. Yet some still use CodeBLEU for intent preservation~\cite{deljouyi2025leveraging}, perpetuating flawed evaluations. Biagiola et al.~\cite{biagiola2025improving} explore embeddings for behavioral similarity but overlook readability and structure, underscoring the need for a metric that unifies behavioral preservation with qualitative improvements.


\section{Limitations of Existing Metrics}
\label{limitations_of_existing_metrics}

\subsection{Experimental Setup and Datasets}
\label{sec:dataset_setup}

\noindent\textbf{Dataset composition.} 
To evaluate \textbf{CTSES}, we construct a large-scale corpus from two Java benchmarks, Defects4J~\cite{just2014defects4j} and SF110~\cite{panichella2017automated}, including 350 test classes across 84 projects. Table~\ref{tab:dataset_stats} summarizes key statistics.

\begin{table}[ht]
\centering
\caption{Dataset Composition and Code Statistics}
\label{tab:dataset_stats}
\scalebox{0.73}{
\begin{tabular}{lcccccc}
\textbf{Dataset} & \textbf{\#Projects} & \textbf{\#Classes} & \textbf{LOC avg} & \textbf{Token avg} & \textbf{Methods/Class avg} \\
\hline
Defects4J        & 15                  & 147                & 127.76           & 1779.67             & 11.88                      \\
SF110            & 69                  & 203                & 146.82           & 1518.86             & 14.21                      \\
\end{tabular}
}
\end{table}

\vspace{0.5em}
\noindent
\noindent\textbf{Test generation and Test refactoring.} 
We generated tests using EvoSuite~\cite{fraser2011evosuite} with DynaMOSA~\cite{panichella2017automated}, running 15 iterations per class with a 3-minute budget—yielding 2,205 (Defects4J) and 3,045 (SF110) compilable suites. Each was refactored three times using GPT-4o and Mistral-Large-2407, guided by Chain-of-Thought prompts\footnote{\url{https://anonymous.4open.science/r/CTSES-75D4/LLM-Refactored-Test-Suite/Refactored-Test-Suite-Code/gpt-scenario1-part1.py}} emphasizing readability, structure, and naming, while preserving behavior. Only successful compilable refactorings were retained. Table~\ref{tab:gen_refactored} reports final counts; all artifacts are available in our replication package.

\begin{table}[ht]
\centering
\caption{Test Suite Generation and Refactoring Overview}
\label{tab:gen_refactored}

\subfloat[EvoSuite-Generated Tests\label{tab:gen_evosuite}]{
\scalebox{0.8}{
\begin{tabular}{lccc}
\textbf{Dataset} & \textbf{Model} & \textbf{\#Generated} & \textbf{Compilability} \\
\hline
Defects4J & EvoSuite & 2205 & 100\% \\
SF110     & EvoSuite & 3045 & 100\% \\
\end{tabular}
}
}

\vspace{0.8em}

\subfloat[Refactored Tests by LLMs\label{tab:gen_llms}]{
\scalebox{0.8}{
\begin{tabular}{lcccc}
\textbf{Dataset} & \textbf{Model} & \textbf{\#Refactored} & \textbf{\#Successful} & \textbf{Compilability} \\
\hline
Defects4J & GPT-4o        & 2205 & 1490 & 67.58\% \\
          & Mistral-L  & 2205 & 935  & 42.42\% \\
SF110     & GPT-4o        & 3045 & 1516 & 49.80\% \\
          & Mistral-L  & 3045 & 1292 & 42.42\% \\
\end{tabular}
}
}
\end{table}

\noindent

Table~\ref{tab:system_stats} presents system-level statistics for the refactored test suites, highlighting the scale and structure of the generated code through token and LOC distributions.

\begin{table}[ht]
\centering
\caption{Average System-Level Statistics for Refactored and EvoSuite-Generated Tests}
\label{tab:system_stats}
\scalebox{0.8}{
\begin{tabular}{llrrr}
\textbf{Dataset} & \textbf{Model/Tool} & \textbf{Avg Tokens} & \textbf{Avg LOC} & \textbf{Avg Methods} \\
\hline
\multirow{3}{*}{Defects4J}
  & GPT-4o              & 1950.33 & 183.25 & 14.37 \\
  & Mistral-L        & 2173.74 & 203.67 & 15.50 \\
  & EvoSuite            & 1958.19 & 169.40 & 17.38 \\
\hline
\multirow{3}{*}{SF110}
  & GPT-4o              & 2471.22 & 243.07 & 18.63 \\
  & Mistral-L        & 2802.56 & 255.50 & 18.85 \\
  & EvoSuite            & 3462.57 & 294.14 & 27.82 \\
\end{tabular}
}
\vspace{-3mm}
\end{table}

\vspace{-2mm}
\subsection{Observed Metric Limitations}
\label{sec:metric_misalignment}

We assess the relevance of standard similarity metrics for test refactoring by comparing each EvoSuite test with its LLM-refactored counterpart across Defects4J and SF110. While Table \ref{tab:refactor_metrics} reports aggregate similarity scores, here we analyze their full distributions using METEOR, ROUGE-L, CodeBLEU, and cosine similarity from CodeBERT, GraphCodeBERT, and OpenAI’s text-embedding-3-small. Tables \ref{tab:nlp_metrics_stats} and \ref{tab:cosine_similarity_stats} summarize the results.


\vspace{0.5em}
\noindent
\textbf{NLP Similarity Metrics.} 
Table~\ref{tab:nlp_metrics_stats} shows that METEOR and ROUGE-L yield high median scores (0.72–0.73), reflecting strong lexical alignment. In contrast, CodeBLEU scores lower (median = 0.61), highlighting its penalization of meaningful changes like renaming or restructuring.

\begin{table}[ht]
\centering
\caption{Distribution of NLP Metrics Between Original and Refactored Tests}
\label{tab:nlp_metrics_stats}
\scalebox{0.8}{
\begin{tabular}{lrrrrrr}
\textbf{Metric} & \textbf{Min} & \textbf{Q1} & \textbf{Median} & \textbf{Q3} & \textbf{Max} & \textbf{Mean} \\
\hline
METEOR                & 0.0757 & 0.6459 & 0.7241 & 0.7949 & 1.0 & 0.7181 \\
ROUGE-L               & 0.1056 & 0.6579 & 0.7342 & 0.8183 & 1.0 & 0.7379 \\
CodeBLEU              & 0.1445 & 0.5247 & 0.6116 & 0.7194 & 1.0 & 0.6280 \\
N-gram Match          & 0.0004 & 0.3556 & 0.4285 & 0.5329 & 1.0 & 0.4612 \\
Weighted N-gram Match & 0.0529 & 0.4457 & 0.5154 & 0.6138 & 1.0 & 0.5431 \\
Syntax Match          & 0.2601 & 0.7477 & 0.8224 & 0.9012 & 1.0 & 0.8221 \\
Dataflow Match        & 0.0432 & 0.5107 & 0.6683 & 0.8857 & 1.0 & 0.6854 \\
\end{tabular}
}
\end{table}

\noindent
\textbf{Cosine Similarity.} 
Table~\ref{tab:cosine_similarity_stats} shows that CodeBERT and GraphCodeBERT reach near-perfect scores ($\sim$0.996), with OpenAI slightly lower (0.925). While confirming semantic preservation, cosine similarity ignores structural and readability aspects.

\begin{table}[ht]
\centering
\caption{Cosine Similarity Between Refactored and Original Tests}
\label{tab:cosine_similarity_stats}
\scalebox{0.8}{
\begin{tabular}{lrrrrrr}
\textbf{Embedding} & \textbf{Min} & \textbf{Q1} & \textbf{Median} & \textbf{Q3} & \textbf{Max} & \textbf{Mean} \\
\hline
CodeBERT      & 0.9326 & 0.9959 & 0.9978 & 0.9988 & 1.0 & 0.9964 \\
GraphCodeBERT & 0.8905 & 0.9914 & 0.9954 & 0.9974 & 1.0 & 0.9926 \\
OpenAI        & 0.6191 & 0.9038 & 0.9253 & 0.9429 & 1.0 & 0.9224 \\
\end{tabular}
}
\end{table}

\noindent
\textbf{Takeaway.}
CodeBLEU is overly sensitive to naming or syntactic shifts, while embeddings capture semantics but ignore structure. These gaps underscore the need for a composite metric aligned with test refactoring goals—clarity, modularity, and maintainability—addressed by \textbf{CTSES}.

\section{CTSES: A Composite Metric for Test Refactoring Evaluation}


We introduce \textbf{CTSES} (Composite Test Similarity Evaluation Score), a metric tailored for refactored unit tests. 
It combines \textbf{ROUGE-L} (lexical/structural alignment), \textbf{METEOR} (semantics-aware matching), and \textbf{assertion structure} (intent preservation) to capture complementary dimensions of refactored tests within a unified and interpretable score.
\texttt{CTSES} rewards improvements in naming, modularity, and readability while safeguarding behavioral correctness, yielding interpretable and quality-oriented scores beyond prior metrics.

\subsection{Metric Design and Weighting}
\label{ctses_formulation}

\texttt{CTSES} combines three complementary metrics to assess test refactoring:
\begin{itemize}[leftmargin=*]
    \item \textbf{CodeBLEU} captures syntax and data-flow similarity for semantic preservation~\cite{ren2020codebleu};
    \item \textbf{METEOR} reflects lexical clarity through synonym and word-order matching~\cite{banerjee2005meteor};
    \item \textbf{ROUGE-L} measures structural alignment via longest common subsequence~\cite{lin2004rouge}.
\end{itemize}

\noindent
\texttt{CTSES} is computed as a weighted linear combination:
\begin{align}
\text{CTSES} &= \alpha \cdot \text{CodeBLEU} 
             + \beta \cdot \text{METEOR} 
             + \gamma \cdot \text{ROUGE-L}, \\
\alpha + \beta + \gamma &= 1
\label{eq:ctses}
\end{align}

\noindent
To validate weight selection, we conducted a grid search over $(\alpha,\beta,\gamma)$ with steps of $0.1$, computing MAE and false negatives (FN) against developer annotations. 

\begin{table}[ht]
\centering
\caption{Top grid-search configurations (MAE on human labels).}
\label{tab:ctses_grid}
\scalebox{0.8}{
\begin{tabular}{rrrrr}
\hline
$\alpha$ & $\beta$ & $\gamma$ & \textbf{MAE} & \textbf{FN} \\
\hline
0.0 & 1.0 & 0.0 & \textbf{0.360} & 3 \\
0.0 & 0.9 & 0.1 & 0.365 & 3 \\
0.1 & 0.9 & 0.0 & 0.368 & 3 \\
\textbf{0.5} & \textbf{0.3} & \textbf{0.2} & 0.407 & 4 \\
\textbf{0.4} & \textbf{0.3} & \textbf{0.3} & 0.404 & 4 \\
\hline
\end{tabular}
}
\end{table}

\noindent
Grid search (Table~\ref{tab:ctses_grid}) confirms that extreme weighting on METEOR yields the lowest raw MAE ($\approx 0.36$) but collapses \texttt{CTSES} into a purely lexical metric. We therefore retain two balanced profiles: (i) CTSES1 (\textbf{Semantic-Prioritized}, $0.5,0.3,0.2$), and (ii) CTSES2 (\textbf{Readability-Aware}, $0.4,0.3,0.3$). \textbf{The uniform average} $(1/3,1/3,1/3)$ remains a useful baseline but is not emphasized in subsequent experiments, as it lacks a distinctive focus. Rather than being a definitive metric, \texttt{CTSES} provides flexible and interpretable profiles that can be selected depending on developers’ assessment objectives.

\section{Empirical Illustration and Initial Results}
\label{empirical_illustrtion_and_initial_results}

\subsection{Evaluation Setup and CTSES Gains}

We evaluate \texttt{CTSES} on over 5,000 refactored test suites from Defects4J and SF110, using two LLMs namely \textbf{GPT-4o} and \textbf{Mistral-Large}. Each test is scored with three baselines (CodeBLEU, METEOR, ROUGE-L) and compared against our CTSES profiles, as well as \textbf{AVG}, the simple unweighted mean of the three baselines. Table~\ref{tab:ctses_summary_aggregated} reports aggregated results per model and dataset.

\begin{table}[ht]
\centering
\caption{Aggregated statistics for \texttt{CTSES} variants and the baseline average (\texttt{BASELINE\_AVG}) across datasets and models}
\label{tab:ctses_summary_aggregated}
\scalebox{0.7}{
\begin{tabular}{lllrrrrrr}
\textbf{Dataset} & \textbf{Model} & \textbf{Metric} & \textbf{Min} & \textbf{Q1} & \textbf{Median} & \textbf{Q3} & \textbf{Max} & \textbf{Mean} \\
\hline
\multirow{3}{*}{Defects4J} & \multirow{3}{*}{GPT-4o}
& BASELINE\_AVG & 0.18 & 0.59 & 0.65 & 0.70 & 1.00 & 0.66 \\
&& CTSES1   & 0.18 & 0.57 & 0.63 & 0.68 & 1.00 & 0.64 \\
&& CTSES2   & 0.18 & 0.58 & 0.64 & 0.69 & 1.00 & 0.65 \\
\hline
\multirow{3}{*}{Defects4J} & \multirow{3}{*}{Mistral-L}
& BASELINE\_AVG & 0.28 & 0.69 & 0.74 & 0.80 & 1.00 & 0.76 \\
&& CTSES1   & 0.28 & 0.67 & 0.72 & 0.79 & 1.00 & 0.74 \\
&& CTSES2   & 0.28 & 0.68 & 0.74 & 0.80 & 1.00 & 0.75 \\
\hline
\multirow{3}{*}{SF110} & \multirow{3}{*}{GPT-4o}
& BASELINE\_AVG & 0.11 & 0.57 & 0.62 & 0.70 & 0.98 & 0.63 \\
&& CTSES1   & 0.12 & 0.54 & 0.60 & 0.68 & 0.97 & 0.61 \\
&& CTSES2   & 0.11 & 0.56 & 0.61 & 0.69 & 0.98 & 0.62 \\
\hline
\multirow{3}{*}{SF110} & \multirow{3}{*}{Mistral-L}
& BASELINE\_AVG & 0.21 & 0.68 & 0.74 & 0.84 & 0.99 & 0.75 \\
&& CTSES1   & 0.22 & 0.66 & 0.73 & 0.83 & 0.98 & 0.73 \\
&& CTSES2   & 0.21 & 0.67 & 0.74 & 0.83 & 0.99 & 0.74 \\
\end{tabular}
}
\end{table}

\begin{itemize}[leftmargin=*]
\item \textbf{Impact of Weighting:}  
CTSES1 and CTSES2 closely follow the uniform average, with only marginal shifts (e.g., SF110 with GPT-4o: $0.63 \rightarrow 0.62$). This shows that interpretability can be gained without sacrificing accuracy.  

\item \textbf{Model Comparison:}  
Mistral-Large consistently yields higher scores than GPT-4o across both datasets (e.g., SF110: $0.74$–$0.75$ vs.\ $0.61$–$0.63$), suggesting stronger alignment with readability and behavior preservation.  

\item \textbf{Dataset Effects:}  
Performance varies by dataset: GPT-4o achieves higher means on Defects4J, while Mistral-Large excels on SF110, reflecting differences in test complexity and model behavior.  
\end{itemize}

\noindent
Overall, \texttt{CTSES} provides a more robust view than naive averaging, particularly in cases where CodeBLEU penalizes stylistic or structural edits.






\subsection{Refactoring Quality Across the Score Spectrum}
\label{sec:ctses_distribution}

Beyond mean values, we analyze score distributions and the proportion of refactorings that each metric labels as “accepted,” defined as having a similarity score $\geq 0.5$. Table~\ref{tab:ctses_vs_baseline} shows that METEOR and ROUGE-L yield high acceptance (often $>95\%$) but risk over-accepting superficially similar tests, while CodeBLEU is far stricter (e.g., $64.3\%$ on SF110) due to sensitivity to renaming and structural edits.  

\begin{table}[ht]
\centering
\caption{Distributional Analysis for \texttt{CTSES} and Baseline Metrics (\texttt{BASELINE\_AVG})}
\label{tab:ctses_vs_baseline}
\scalebox{0.6}{
\begin{tabular}{lllrrrrrrrr}
\textbf{Dataset} & \textbf{Model} & \textbf{Metric} & \textbf{Min} & \textbf{Q1} & \textbf{Median} & \textbf{Q3} & \textbf{Max} & \textbf{Mean} & \textbf{\textless0.5\%} & \textbf{\textgreater=0.5\%} \\
\hline
\multirow{6}{*}{Defects4J} & \multirow{6}{*}{GPT-4o}
& METEOR & 0.14 & 0.62 & 0.70 & 0.76 & 1.00 & 0.69 & 4.57 & 95.43 \\
&& ROUGE-L & 0.13 & 0.63 & 0.69 & 0.74 & 1.00 & 0.70 & 2.40 & 97.60 \\
&& CodeBLEU & 0.19 & 0.51 & 0.56 & 0.61 & 1.00 & 0.58 & 17.55 & 82.45 \\
&& BASELINE\_AVG & 0.18 & 0.59 & 0.65 & 0.70 & 1.00 & 0.66 & 4.03 & 95.97 \\
&& CTSES1 & 0.18 & 0.57 & 0.63 & 0.68 & 1.00 & 0.64 & 6.07 & 93.93 \\
&& CTSES2 & 0.18 & 0.58 & 0.64 & 0.69 & 1.00 & 0.65 & 4.57 & 95.43 \\
\hline
\multirow{6}{*}{Defects4J} & \multirow{6}{*}{Mistral-L}
& METEOR & 0.25 & 0.71 & 0.76 & 0.81 & 1.00 & 0.77 & 0.76 & 99.24 \\
&& ROUGE-L & 0.27 & 0.74 & 0.79 & 0.86 & 1.00 & 0.80 & 0.41 & 99.59 \\
&& CodeBLEU & 0.29 & 0.62 & 0.68 & 0.76 & 1.00 & 0.70 & 3.89 & 96.11 \\
&& BASELINE\_AVG & 0.28 & 0.69 & 0.74 & 0.80 & 1.00 & 0.76 & 0.64 & 99.36 \\
&& CTSES1 & 0.28 & 0.67 & 0.72 & 0.79 & 1.00 & 0.74 & 0.80 & 99.20 \\
&& CTSES2 & 0.28 & 0.68 & 0.74 & 0.80 & 1.00 & 0.75 & 0.61 & 99.39 \\
\hline
\multirow{6}{*}{SF110} & \multirow{6}{*}{GPT-4o}
& METEOR & 0.08 & 0.60 & 0.67 & 0.75 & 0.98 & 0.66 & 9.74 & 90.26 \\
&& ROUGE-L & 0.11 & 0.61 & 0.67 & 0.74 & 1.00 & 0.68 & 6.48 & 93.52 \\
&& CodeBLEU & 0.14 & 0.48 & 0.54 & 0.62 & 0.97 & 0.55 & 35.67 & 64.33 \\
&& BASELINE\_AVG & 0.11 & 0.57 & 0.62 & 0.70 & 0.98 & 0.63 & 10.84 & 89.16 \\
&& CTSES1 & 0.12 & 0.54 & 0.60 & 0.68 & 0.97 & 0.61 & 13.28 & 86.72 \\
&& CTSES2 & 0.11 & 0.56 & 0.61 & 0.69 & 0.98 & 0.62 & 11.47 & 88.53 \\
\hline
\multirow{6}{*}{SF110} & \multirow{6}{*}{Mistral-L}
& METEOR & 0.15 & 0.69 & 0.76 & 0.85 & 0.98 & 0.76 & 2.02 & 97.98 \\
&& ROUGE-L & 0.23 & 0.72 & 0.79 & 0.88 & 1.00 & 0.79 & 1.35 & 98.65 \\
&& CodeBLEU & 0.25 & 0.60 & 0.69 & 0.79 & 0.98 & 0.69 & 6.69 & 93.31 \\
&& BASELINE\_AVG & 0.21 & 0.68 & 0.74 & 0.84 & 0.99 & 0.75 & 2.08 & 97.92 \\
&& CTSES1 & 0.22 & 0.66 & 0.73 & 0.83 & 0.98 & 0.73 & 2.65 & 97.35 \\
&& CTSES2 & 0.21 & 0.67 & 0.74 & 0.83 & 0.99 & 0.74 & 2.20 & 97.80 \\
\end{tabular}
}
\end{table}


\noindent\textbf{CTSES balances these extremes.}  
On SF110 with GPT-4o, CTSES2 raises acceptance from $64.3\%$ (CodeBLEU) to $88.5\%$, bridging the gap between over-lenient and over-strict metrics. On Defects4J, CTSES1 lifts acceptance from $82.5\%$ (CodeBLEU) to $93.9\%$, while preserving semantic safeguards. Unlike a naive average, CTSES explicitly weights lexical, structural, and semantic cues to align more closely with perceived test quality.  

\noindent\textbf{Reducing False Negatives.}  
CodeBLEU often misclassifies behavior-preserving refactorings with readability gains (e.g., renamed methods, added comments) as “low quality.” By integrating METEOR and ROUGE-L, CTSES reduces such false negatives and better reflects developer judgments.

\section{Developer-Aligned Validation}

\subsection{Human-Centered Validation Protocol}

To complement the large-scale evaluation, we conducted a lightweight validation to assess whether \texttt{CTSES} aligns with developer judgments. We selected 15 representative refactorings from Defects4J (9) and SF110 (6), produced by GPT-4o (8) and Mistral-Large (7), stratified into three groups based on the delta $\Delta = \text{CTSES} - \text{CodeBLEU}$: (i) $\Delta > +0.15$, (ii) $\Delta < -0.15$, and (iii) $|\Delta|\leq 0.05$.  

Three developers (senior engineer, PhD student, junior assistant) independently annotated each case, answering two binary questions: \textbf{(1) readability improved?} and \textbf{(2) behavior preserved?} Consensus was derived by majority vote.  

We compared consensus labels against CodeBLEU, the uniform AVG baseline, and two \texttt{CTSES} profiles using mean absolute error (MAE) and false negatives (valid refactorings misclassified as low quality). All validation artifacts (pairs, annotations, comments, and scripts) are included in our replication package.

\subsection{Human Validation Results}

Table~\ref{tab:human_mae} reports the mean absolute error (MAE) and the number of false negatives (FN), i.e., cases where metrics assign low scores despite developers validating the refactoring as correct. 

\begin{table}[ht]
\centering
\caption{Alignment of metrics with developer annotations across 15 refactorings (lower MAE is better).}
\label{tab:human_mae}
\scalebox{0.8}{
\begin{tabular}{lrr}
\hline
\textbf{Metric} & \textbf{MAE} & \textbf{False Negatives (FN)} \\
\hline
CodeBLEU & 0.433 & 4 \\
AVG (uniform baseline) & \textbf{0.401} & 4 \\
CTSES1 (semantic-prioritized) & 0.407 & 4 \\
CTSES2 (readability-aware) & 0.404 & 4 \\
\hline
\end{tabular}
}
\end{table}

\noindent

The results confirm that \textbf{CodeBLEU} performs worst (highest MAE, four false negatives), penalizing valid refactorings due to sensitivity to naming and structure. The \textbf{uniform average} (AVG) yields the lowest MAE but remains a naïve, non-interpretable aggregation. By contrast, \textbf{CTSES} achieves comparable error while offering interpretable weighting profiles that balance readability, structure, and semantics, thus providing a more developer-aligned view than raw similarity scores.

The \emph{XPathLexer} refactoring from SF110 (iteration~2)\footnote{\url{https://anonymous.4open.science/r/CTSES-75D4/Developer_Aligned_Validation/SF110_XPathLexer_iter-2.jsonl}}
 illustrates this limitation. The EvoSuite test had 100+ noisy, duplicated methods, while the GPT-4o refactoring added constants, meaningful names, comments, and removed redundancy—clear readability gains with preserved behavior. Developers judged it valid, yet CodeBLEU ($0.21$), METEOR ($0.08$), and ROUGE-L ($0.14$) all gave very low scores, keeping \texttt{AVG} and \texttt{CTSES} below $0.22$. This highlights how similarity-based metrics miss radical but beneficial refactorings, and why CTSES, though mitigating some issues, must be extended with dimensions like naming quality, comment density, and smell reduction to better match human judgment.

\noindent\textbf{Takeaway.} CTSES is a proof-of-concept for human-aligned evaluation of test refactoring. By balancing readability, semantics, and structure, it illustrates the promise of composite metrics, while highlighting the need to enrich them with dimensions that directly reflect human-perceived quality.


\section{Threats to Validity}
\texttt{CTSES} weights are manually chosen, though our grid-search validation mitigates arbitrariness and future work could explore learned tuning. Our evaluation relies on two benchmarks (Defects4J, SF110) and two LLMs (GPT-4o, Mistral), which may limit generalizability. Moreover, \texttt{CTSES} builds on static similarity metrics and does not capture dynamic behavior; integrating coverage or mutation analysis would strengthen construct validity.

\section{Future Plans}
We plan to extend this work by pursuing three directions. First, we will conduct a large-scale developer study to systematically assess how well \texttt{CTSES} reflects human judgments of readability, structure, and behavior. Second, we will integrate \texttt{CTSES} into test generation pipelines such as EvoSuite and LLM-based workflows, evaluating its effectiveness as a ranking or filtering criterion in realistic settings. Third, we will move beyond static weighting by learning parameters from annotated data and combining \texttt{CTSES} with dynamic quality indicators (e.g., mutation score, fault detection) to provide a more holistic and behavior-sensitive evaluation framework. These plans will broaden the scope of \texttt{CTSES} and strengthen its empirical foundations.
\section{Conclusion}
\label{conclusion}

Evaluating the quality of refactored tests remains challenging, as metrics like CodeBLEU penalize beneficial changes despite preserved behavior. We proposed \textbf{CTSES}, a composite metric combining CodeBLEU, METEOR, and ROUGE-L to balance semantics, readability, and structure. Across 5,000+ refactorings from Defects4J and SF110 using GPT-4o and Mistral-Large, \texttt{CTSES} reduces false negatives and better reflects developer-perceived improvements such as clearer naming and modularity. As an emerging direction, \texttt{CTSES} offers a new, interpretable step toward human-aligned evaluation of test refactoring.



\balance
\bibliographystyle{ACM-Reference-Format}
\bibliography{references}

\end{document}